\documentclass[a4paper,12pt]{article}
\usepackage{a4wide}
\usepackage{graphicx,subfigure,hhline}
\usepackage[colorlinks=true,linkcolor=blue,citecolor=red]{hyperref}
\usepackage{xcolor}
\usepackage{cite}
\usepackage{amssymb,amsthm,empheq,bbold}
\usepackage[inline]{enumitem}
\usepackage[ruled,vlined,linesnumbered]{algorithm2e}
\usepackage{caption} 
\newtheorem{theorem}{Theorem} 

\newtheorem{remark}{Remark}

\usepackage{subfigure}

\definecolor{amber}{rgb}{1.0, 0.49, 0.0}

\allowdisplaybreaks

\begin{document}
%	\begin{linenumbers}
	\title{A nonconservative kinetic framework with logistic growth for modeling the coexistence in a multi-species ecological system}
	\author{Marco Menale$^1$, Carmelo Filippo Munaf\`o$^2$, Francesco Oliveri$^2$,\\[1em]
	$^{1}${\footnotesize Dipartimento di Matematica e Applicazioni ``R. Caccioppoli", Federico II} \\ {\footnotesize Via Cintia, Monte S. Angelo I-80126 Napoli, Italy}\\
	$^2${\footnotesize Department of Mathematical and Computer Sciences, Physical Sciences and Earth Sciences,} \\ {\footnotesize University of Messina, Viale F. Stagno d'Alcontres 31, 98166, Messina Italy}\\
\\{\footnotesize 
	(*corresponding author)}
	\\[0.5em]
	{\footnotesize (*) marco.menale@unina.it} \\
        {\footnotesize carmelofilippo.munafo@unime.it} \\
        {\footnotesize francesco.oliveri@unime.it}}	
	
    \date{}
    \maketitle
    
 \begin{abstract}
Kinetic theory frameworks are widely used for modeling stochastic interacting systems, where the evolution primarily depends on binary interactions. Recently, in this framework the action of the external force field has been introduction in order to gain a more realistic picture of some phenomena. In this paper, we introduce nonconservative kinetic equations where a particular shape external force field acts on the overall system. Then, this framework is used in an ecological context for modeling the evolution of a system composed of two species interacting with a prey--predator mechanism. The linear stability analysis concerned with the coexistence equilibrium point is provided, and a case where a Hopf bifurcations occurs is discussed. Finally, some relevant scenarios are numerically simulated.
\end{abstract}

\noindent{\bf Keywords:} Kinetic theory; Differential Equations; Nonconservative dynamics; Stability; Ecological systems.

\smallskip

\noindent{\bf MSC:} 82B40, 82C22, 34A12, 37N25. 

\maketitle

\section{Introduction}

The kinetic theory has been widely used in the last years to model the evolution of interacting systems in various fields: opinion dynamics \cite{bertotti2008discrete,pareschi2019hydrodynamic, toscani2006kinetic, toscani2018opinion}, economics \cite{bertotti2012exploiting, letizia2016economic, bertotti2017stochastic}, psychology \cite{bellomo2008complexity}, traffic-flow problems \cite{albi2019vehicular, puppo2017analysis}, epidemiology \cite{albi2022kinetic, della2023sir, loy2021viral}, socio-economic impact of a disease outbreak \cite{bernardi2022effects,dimarco2020wealth}, and behavioral epidemiology \cite{della2023intransigent}, among others. Moreover, recently (see, for instance, \cite{toscani2023kinetic,menale2024kinetic}) the kinetic modeling tools have been extended to ecological problems. In particular, in \cite{toscani2023kinetic}, a kinetic derivation of a Lotka-Volterra model is provided by using Fokker-Planck equation, whereas in \cite{menale2024kinetic}, discrete kinetic equations are used for modeling an ecological system composed of prey and predators characterized by their expertise level. The kinetic theory has its great advantage in the multiscale perspective (see \cite{bellomo2008modeling,bellomo2014multiscale}). 

In this paper, a kinetic model for an ecological system accounting for population growth is proposed. In particular, a binary and stochastic interacting system is considered at three different scales or levels: \emph{microscopic}, \emph{mesoscopic}, and \emph{macroscopic}. At microscopic scale, the stochastical binary interactions between pairs of individuals, also called \emph{agents} or \emph{particles}, are considered; the microscopic state of each individual is characterized by one or more real variables, that can be either continuous or discrete. At the mesoscopic level, representing the statistical level of the modeling approach, we are no longer concerned with binary interactions, since we regard the overall distributions in the system. This will be the main focus of the analysis carried out in this paper. The specific analytic form of the evolution equations at the mesoscopic level depends on the particular microscopic framework considered. Indeed, one can have: ordinary differential equations, partial differential equations, integro-differential equations. Hereafter, for the purposes of this paper, we will limit ourselves to consider a system of nonlinear ordinary differential equations, since we assume a discrete microscopic variable. Finally, at the macroscopic level, the global properties of the system are obtained.  It is worth stressing that a multiscale approach allows us to shed light on various aspects of the evolution of the phenomenon at hand. Generally, in the conservative case, \emph{i.e.}, when the overall amount of agents does not change during the evolution of the system, there exists a unique and positive solution of the system. Nevertheless, the conservative structure may be too restrictive in some cases if a realistic picture of a phenomenon is required. Therefore, in this paper we will develop a nonconservative kinetic framework, whereupon the overall amount of agents will not be constant during the time evolution; this aspect is of interest as highlighted in some other papers (see, for instance, \cite{bellouquid2011kinetic, bianca2017modeling,bianca2024decade,bianca2024discrete}, and references therein).

The kinetic model developed in this paper originates from an ecological problem, involving $n$ different populations, or species. Generally, in the last decades, the ecological problems have been of an increasing interest for the scientific community,  as shown by the wide literature on these topics \cite{ajraldi2011modeling,chen2023dynamic,chowdhury2022canards,laurie2020herding,liang2022nonlocal,malchow2007spatiotemporal,rosenzweig1963graphical,sun2022dynamic,sun2022impacts}. Ecological models have found widespread use in various fields, such as analyzing and conserving interacting species \cite{oliveira2010modelling}, understanding the spread of infectious diseases (refer to \cite{venturino2016ecoepidemiology} for more details), managing crop pest populations \cite{bhattacharyya2006pest,jana2013mathematical}, and controlling harmful algal blooms in aquatic environments, which threaten both the fishing industry and tourism \cite{chattopadhayay2002toxin}. By using the approach in \cite{bellomo2009complexity}, the overall ecological system is divided into functional subsystems such that each of them represents a certain species. The individuals of populations are assumed to stochastically pairwise interact at the microscopic scale. Then, the kinetic equations provide the evolution of each functional subsystem at the mesoscopic level. As far as the nonconservative events are concerned, we will consider growth phenomena of the involved populations both from an ecological and mathematical point of view. From an ecological perspective, we aim at modeling the impact of a logistic growth of each population on all the species. From a mathematical point of view, we include this effect by introducing, at the mesoscopic level, an \emph{external force field} in the system of kinetic equations. This aspect has been recently investigated in an ecological \cite{carbonaro2023nonconservative} as well as an epidemiological application \cite{menale2023kinetic}. In the latter papers, the external force field has the role of modeling the impact of the external environment on each functional subsystem. Notice that the introduction of nonconservative terms in a kinetic system may lead to the loss of certain analytical properties, such as, for example, the boundedness of solutions (as shown in the continuous case in \cite{arlotti1999jager}). Therefore, some further analytical assumptions are required in order to preserve the well posedness of the problem because the solution is required to be positive and bounded. Then, some first stability results are obtained. In particular, among the equilibria provided by the system, we will focus our attention on the coexistence equilibrium point, that is the equilibrium point that ensures the survival of both populations (see for details \cite{chattopadhyay2008patchy}, and references therein).

The paper is organized as follows. In Section~\ref{secmodel}, the nonconservative kinetic model with growth term, expressed by an external force field, is introduced and discussed. Section~\ref{sececolg} provides the ecological interpretation of the model; in particular, the meaning of parameters and external force field is specified with respect to the particular case of an ecological system made of $2$ populations subjected to logistic growth effects. In Section~\ref{seccoex}, a first stability analysis is provided, and in Subsections~\ref{subsecfirts}--\ref{subsecfou}, four different scenarios are analyzed by assigning specific values to the parameters of the system. The analysis is restricted to  the coexistence equilibrium point and its stability properties. In particular, a first Hopf bifurcation is shown in Theorem~\ref{teo_fourth_scenario}. Finally, Section~\ref{SecConcl} contains our conclusions.

\section{The kinetic model}\label{secmodel}

This Section is devoted to the derivation of a nonconservative kinetic framework subjected to a specific external force field. Specifically, we provide a general derivation, without yet addressing the specific ecological application of this paper. 

First, let us consider a stochastically interacting systems, composed of \emph{particles}, also called \emph{agents}. The overall system is divided into $n\in \mathbb{N}$ \emph{functional subsystems}, such that agents belonging to the same functional subsystem share the same strategy \cite{bellomo2009complexity}. Specifically, the meaning of each functional subsystem depends on the particular applications taken into account.

Thus, in the kinetic approach (see \cite{tosin2019kinetic}, and references therein) the evolution of the system is characterized at different scales. At the \emph{microscopic level}, the system is described by using real variables $u_i$ $(i=1,\ldots,n)$, called \emph{activity variables}, which attain values in a real discrete subset, \emph{i.e.}, $u_i \in D_i \subseteq \mathbb{R}$. As well as for functional subsystems, the meaning of activity variables depends on the particular application considered; for instance, they may express the viral load \cite{della2023sir}. Hereafter, for the sake of simplicity, but without constraining the modeling approach of this paper, we identify functional subsystems and activity variables $u_i$ $(i=1,\ldots,n)$. The binary and stochastic interaction between pairs of agents is defined by the following quantities:
\begin{itemize}
\item the \emph{transition probability} $B^i_{hk}$ $(i,h,k=1, \ldots, n)$, which describes the probability that an agent of the $h$th functional subsystem after interacting with an agent of the $k$th functional subsystem falls into the $i$th functional subsystem. Since it is a probability, the following constraint has to be satisfied:
\begin{equation}
\label{assump}
\sum_{i=1}^n B^i_{hk}=1, \qquad \forall h,k \in\{1, \ldots, n\}.
\end{equation}
\item the \emph{interaction rate} $\eta_{hk}$ $(h,k \in \{1, \ldots, n\})$, which determines the number of encounters between agents of the $h$th functional subsystem and agents of the $k$th functional subsystem.
\end{itemize}
At the \emph{mesoscopic level}, the state of the $i$th functional subsystem at time $t > 0$ is described by the \emph{distribution function} 
\[
f_i(t):[0,\, T]\rightarrow \mathbb{R}^+,
\] 
which determines the number of agents in the $i$th functional subsystem at time $t>0$. Moreover, the \emph{vector distribution function} related to the overall system writes, for $t>0$, 
\begin{equation*}
\mathbf{f}(t)=\left(f_1(t), f_2(t), \ldots, f_n(t)\right).
\end{equation*}
Thus, the \emph{kinetic equation}  related to the $i$th functional subsystem is  \cite{bertotti2004discrete}
\begin{equation}\label{eqcons}
    \frac{df_i(t)}{dt}=\sum_{h,k=1}^n\eta_{hk}B^i_{hk}f_h(t)f_k(t)-f_i(t)\sum_{k=1}^n\eta_{ik}f_k(t),
    \qquad i=1,\ldots,n,
\end{equation}
to be solved with the nonnegative \emph{initial data} 
\begin{equation}
\label{cauchycons}
\mathbf{f}(0)=\mathbf{f}^0\in \left(\mathbb{R}^+\right)^n.
\end{equation}

Finally, the \emph{macroscopic state} of the overall system is described by introducing the $p$\emph{th-order moment} of the system, for $p \in \mathbb{N}$, say,
\begin{equation*}
\mathbb{E}_p[\mathbf{f}](t):=\sum_{i=1}^nu_i^pf_i(t).
\end{equation*}
From a physical viewpoint, the $0$th-order moment, the $1$st-order moment, and the $2$nd-order moment represent the density, the linear momentum and the global activation energy of the system, respectively. We notice that for the evolution of an ecological system, the $0$th-order moment may be of interest, as it represents the total amount of agents of the system itself. Hereafter, we use the symbol $\rho(t)$ to refer to the $0$th-order moment, say the \emph{global density} of the system.

Due to the constraint~\eqref{assump}, the kinetic framework~\eqref{eqcons} is \emph{conservative}: 
\begin{equation*}
\frac{d\rho(t)}{dt}=0, \qquad \forall t>0.
\end{equation*}
Therefore, the Cauchy problem~\eqref{eqcons}--\eqref{cauchycons} has a unique, bounded and positive solution $\mathbf{f}(t)$, globally in time (for details, see \cite{bertotti2004discrete}). In particular, if the initial data $\mathbf{f}^0$ are such that
\begin{equation*}
\sum_{i=1}^nf_i^0=1,
\end{equation*}
then the solution $\mathbf{f}(t)$ satisfies  
\begin{equation*}
\sum_{i=1}^nf_i(t)=1, \qquad \forall t>0.
\end{equation*}
Then, the solution of the framework~\eqref{eqcons}--\eqref{cauchycons} can be interpreted as a \emph{probability}.

As stated above, we now introduce an \emph{external action} on the overall system. In the system of kinetic equations~\eqref{eqcons}, this action is represented by an \emph{external force field} $\mathbf{F}[\mathbf{f}](t)$ with components
\[
F_i[\mathbf{f}](t):[0,\, T] \rightarrow \mathbb{R}, \qquad i=1,\ldots,n.
\]
Then, the \emph{kinetic equations with the external force field} $\mathbf{F}[\mathbf{f}](t)$ writes (see for details \cite{carbonaro2023nonconservative})
\begin{equation}\label{eqforc}
\frac{df_i(t)}{dt}=\sum_{h,k=1}^n\eta_{hk}B^i_{hk}f_h(t)f_k(t)-f_i(t)\sum_{k=1}^n\eta_{ik}f_k(t)+F_i[\mathbf{f}](t), \qquad i=1,\ldots,n.
\end{equation}
In general, the framework~\eqref{eqforc} is not conservative, and this may cause the loss of positivity and/or boundedness of solution. Indeed, in nonconservative kinetic frameworks, blow-up phenomena may also occur \cite{arlotti1999jager}.

In this paper, we consider a specific analytical form for the external force field $\mathbf{F}[\mathbf{f}](t)$, that is
\begin{equation}\label{shapeforc}
F_i[\mathbf{f}](t)=\sum_{h=1}^n\Gamma_i^h(t)f_h(t)\left(1-f_h(t)\right),
\end{equation}
where $\Gamma^h_i(t)$'s are, at least, continuous functions. Roughly speaking, \eqref{shapeforc} can be interpreted as follows. For each functional subsystem, the action of the external force field depends on all the other functional subsystems, due to coefficients $\Gamma_i^h(t)$; it is worth noting that these coefficients are not the same for all the functional subsystems. These functions are independent of the current state of the related functional subsystem. Nevertheless, for a more realistic description of the evolution of a kinetic system, the action of an external environment on a stochastically binary interacting system has to depend on the current state of the system. Therefore, for these reasons, the term $f_h(t)(1-f_h(t))$ in \eqref{shapeforc} provides the dependence of this action with respect to the current state of the $i$th functional subsystem, and all the others, \emph{i.e.}, for all $h\neq i$. Furthermore, this particular choice aims at mimicking a logistic evolution for each functional subsystem. This latter aspect will become clear below. We remark that the structure~\eqref{shapeforc} is only one of the possible choices for an external force field. Indeed, in \cite{menale2023kinetic}, a different external force field was assumed, without the quadratic nonlinearity~\eqref{shapeforc} related to the distribution functions $f_i$'s.

Therefore, the framework~\eqref{eqforc} now writes as
\begin{equation}\label{eqnewfor}
    \frac{df_i(t)}{dt}=\sum_{h,k=1}^n\eta_{hk}B^i_{hk}f_h(t)f_k(t)-f_i(t)\sum_{k=1}^n\eta_{ik}f_k(t)+\sum_{h=1}^n\Gamma_i^h(t)f_h(t)\left(1-f_h(t)\right).
\end{equation}
It is easy to ascertain that the model~\eqref{eqnewfor} is nonconservative; in fact, one has
\begin{equation}\label{eqdensity}
    \frac{d\rho(t)}{dt}=\sum_{i=1}^n\sum_{h=1}^n\Gamma_i^h(t)f_h(t)\left(1-f_h(t)\right).
\end{equation}
Consequently, the quantity $\rho(t)$ could blow-up, or the positivity of the solution $\mathbf{f}(t)$ of the framework~\eqref{eqnewfor} could not be guaranteed,  even for positive initial data $\mathbf{f}^0\in\left(\mathbb{R}^+\right)^n$. For instance, negative values for the coefficients $\Gamma_i^h(t)$, for some $i,h \in \{1, \ldots, n\}$, could determine the negativity of the solution. Therefore, we can expect only existence and uniqueness of a solution, at most locally in time, as shown by the next result.

\begin{theorem}\label{th1}
Let assume the kinetic framework~\eqref{eqnewfor}. Along with the constraint~\eqref{assump}, suppose that the three following further assumptions are satisfied:
\begin{description}
        \item[(i)] the functions $\Gamma_i^h(t)$ are continuous for all $i,h \in \{1, \ldots, n\}$, and for all $t>0$;
        \item[(ii)] there exists $\eta>0$ such that 
        $$\eta_{hk}\leq \eta,\quad \forall h,k \in \{1, \ldots, n\};$$
        \item [(iii)] the initial data are nonnegative, say 
$$\mathbf{f}^0=\left(f_1^0,\ldots, f^0_n\right)\in \left(\mathbb{R}^+\right)^n.$$
\end{description}
Then, there exists a unique solution $\mathbf{f}(t)=\left(f_1(t),\dots, f_n(t)\right)$ in the time interval $[0,\, t_0]$, that is positive and bounded.
    
\begin{proof}    
Using the same arguments developed in \cite{menale2023kinetic}, one can prove by standard arguments (see, for instance, \cite{coddington2012introduction}) that the right-hand side of the equation~\eqref{eqnewfor} is Lipschitzian, for all $i \in \{1, \ldots, n\}$. Therefore, there exists a unique and bounded solution $\mathbf{f}(t):[0,\, t_0]\to \mathbb{R}^n$, local in time, of the Cauchy problem related to the framework~\eqref{eqnewfor}. In particular, $t_0$ depends on the initial data and the parameters of the system itself.
     
Let us consider the assumption \textbf{(iii)}, \emph{i.e.}, the initial data $\mathbf{f}^0$ are such that each component is strictly positive. Then, there exists, by continuity, a bounded and positive solution, at least in a small time interval $[0,\, t_0]$. Therefore, at least locally in time, we can conclude that there is a time $t_0>0$ such that there exists a unique solution $\mathbf{f}(t)=\left(f_1(t), \ldots, f_n(t))\right)$ of the Cauchy problem related to the system~\eqref{eqnewfor} that is bounded and positive. This concludes the proof.
\end{proof}     
\end{theorem}

\begin{remark}
Theorem~\ref{th1} ensures that a bounded and positive solution $\mathbf{f}(t)$ of the kinetic framework~\eqref{eqnewfor} exists, even if local in time. In order to extend this solution globally in time, the density $\rho(t)$ needs to be bounded, that is there must exist a positive constant $K$ such that, for all $t>0$,
\begin{equation}
0 \le \rho(t)=\rho(0)+\sum_{i=1}^n\sum_{h=1}^n\int_0^t\Gamma_i^h(\tau)f_h(\tau)\left(1-f_h(\tau)\right)\,d\tau<K,
\end{equation}
For instance, if the function $\Gamma_i^h(t)f_h(t)\left(1-f_h(t)\right)$ is integrable on the interval $[0,+\infty[$, for $i,h \in \{1, \ldots, n\}$, then the boundedness of solution ensures that the solution $\mathbf{f}(t)$ can be extended globally in time. 

It is worth stressing that neither the boundedness, nor the positiveness can be easily derived from the kinetic model~\eqref{eqnewfor}.
\end{remark}

\section{The ecological application}
\label{sececolg}

Herafter, we will use the nonconservative kinetic framework~\eqref{eqnewfor} in order to model an ecological system made of two species interacting with a predator-prey mechanism  \cite{lotka2002contribution, volterra1926variazioni, volterra1928variations}. The first two terms of the kinetic system~\eqref{eqnewfor} model the interactions among individuals of the system. On the contrary, the latter term refers to the action of the external environment on each population (see \cite{carbonaro2023nonconservative}). Nevertheless, with respect to the previous study, where the external force field acting on a certain functional subsystem depends only on the functional subsystem itself, the structure~\eqref{shapeforc} of the external force field assumed in this paper is slightly different. Indeed, the action of the external force field $F_i[\mathbf{f}]$ acting on the $i$th functional subsystem depends on the current state of all functional subsystems. It is worth noting that the structure $f_h(t)(1-f_h(t))$ is logistic (see for details \cite{mandal2020nonautonomous,tsoularis2002analysis}). Therefore, the evolution of each population depends, besides binary interactions, on the external environment that acts according to the growth of all involved populations. 

There are two functional subsystems: $f_1(t)$ represents the distribution function of the first population (predators), whereas $f_2(t)$ stands for the distribution function of the second population (prey). 

For the sake of simplicity, but as a first approximation, the coefficients $\Gamma_i^h(t)$, for $i,h\in \{1,\, 2\}$, are assumed to be time-independent.

Thus, the nonconservative kinetic system~\eqref{eqnewfor} specialize to the following pair of nonlinear ordinary differential equations:
\[ 
\begin{aligned}
& \frac{df_1}{dt} = - \eta_{11}B^2_{11}f_1^2 - \eta_{12}B^2_{12}f_1 f_2 + \eta_{21}B^1_{21}f_1f_2 + \eta_{22}B^1_{22}f_2^2 + \Gamma_1^1 f_1\left(1-f_1\right) + \Gamma_1^2 f_2\left(1-f_2\right), 
\\
& \frac{df_2}{dt} = \eta_{11}B^2_{11}f_1^2 + \eta_{12}B^2_{12}f_1f_2-\eta_{21}B^1_{21}f_1f_2 - \eta_{22}B^1_{22}f_2^2 + \Gamma_2^1 f_1\left(1-f_1\right) + \Gamma_2^2 f_2\left(1-f_2\right).
\end{aligned}    
\]
Hereafter, we further suppose that $\eta_{11}=\eta_{22}=0$, that is we assume that the evolution of each population is affected by the stochastic binary interaction among pairs of individuals of different species, besides the action of the external environment, weighted on the growth of both populations. Furthermore, we choose 
$\eta_{12}=\eta_{21}\equiv \eta>0$,
\emph{i.e.}, we take symmetric interaction rates: \emph{i.e.}, if a predator encounters a prey, it implies that the prey has also encountered the predator. Finally, introducing the real parameter $\alpha$ such that
\begin{equation}\label{eqalpha}
\alpha:= \eta\left(B_{12}^2-B_{21}^1\right).
\end{equation}
we are led to the system
\begin{equation}\label{sys2}
    \begin{aligned}
	& \frac{df_1}{dt} = -\alpha f_1f_2 + \Gamma_1^1 f_1\left(1-f_1\right) + \Gamma_1^2 f_2\left(1-f_2\right), \\
        & \frac{df_2}{dt} = \alpha f_1f_2 + \Gamma_2^1 f_1\left(1-f_1\right) + \Gamma_2^2f_2\left(1-f_2\right),
    \end{aligned}    
\end{equation}
to be solved with the initial condition 
$$\mathbf{f}^0=\left(f^0_1,\, f^0_2\right)\in (\mathbb{R}^+)^2,$$
where $f^0_1$ and $f_0^2$ represent the number of predators and prey at the initial time, respectively. As above remarked, the Cauchy problem admits a positive solution in a time interval $[0,\, t_0]$. Hereafter, we will focus only on cases where $f_1(t),\, f_2(t)>0$, for $t\geq 0$. Furthermore,  also vanishing  values of the solution are not of interest in this paper, so that we exclude the cases where at least one of the two populations goes extinct. 

\begin{remark}
In general, when a logistic growth for a population is assumed in a model, then a carrying capacity is introduced, \cite{mandal2020nonautonomous,tsoularis2002analysis},  in order to fix the maximum size that the population can attain in a certain environment. Roughly speaking, in the kinetic model~\eqref{sys2}, the carrying capability for each population is implicitly assumed equal to $1$. In a more general framework, we can introduce two different carrying capacities $k_1$, $k_2 >0$ for the first and second species, respectively. Then, the model~\eqref{sys2} rewrites as
\begin{equation}\label{syscap}
    \begin{aligned}
	& \frac{df_1}{dt} = -\alpha f_1 f_2 + \Gamma_1^1\,f_1\left(1-\frac{f_1}{k_1}\right) + \Gamma_1^2 \,f_2\left(1-\frac{f_2}{k_2}\right), \\
        & \frac{df_2}{dt} = \alpha f_1 f_2 + \Gamma_2^1\,f_1\left(1-\frac{f_1}{k_1}\right) + \Gamma_2^2\,f_2\left(1-\frac{f_2}{k_2}\right).
    \end{aligned}    
\end{equation}
Let us set
\[
F_1:=\frac{f_1}{k_1},\qquad F_2:=\frac{f_2}{k_2},
\]
whence  the system~\eqref{syscap} reduces to
\begin{equation}\label{syscap_1}
    \begin{aligned}
        & \frac{dF_1}{dt} = -\alpha k_2 F_1 F_2 + \Gamma_1^1\,F_1\left(1-F_1\right) + \frac{k_2}{k_1}\,\Gamma_1^2 \,F_2\left(1-F_2\right), \\
        & \frac{dF_2}{dt} = \alpha k_1 F_1 F_2 + \frac{k_1}{k_2}\,\Gamma_2^1\,F_1\left(1-F_1\right) + \Gamma_2^2\,F_2\left(1-F_2\right).
    \end{aligned}    
\end{equation}
The analytical structure is not different if compared to the previous system~\eqref{sys2}, apart the form of the parameters. Indeed, if the following quantities are introduced 
\begin{equation*}
   \tilde{\alpha}_1 :=  \alpha k_1,\quad \tilde{\alpha}_2 :=  \alpha k_2,  \quad \tilde{\Gamma}_1^2 := \frac{k_2}{k_1}\,\Gamma_1^2, \quad \tilde{\Gamma}_2^1 :=\frac{k_1}{k_2}\,\Gamma_2^1,
\end{equation*}
then, the system~\eqref{syscap_1} rewrites as
\begin{equation}\label{syscap_2}
    \begin{aligned}
        & \frac{dF_1}{dt} = - \tilde{\alpha}_2 F_1 F_2 + \Gamma_1^1\,F_1\left(1-F_1\right) + \tilde{\Gamma}_1^2 \,F_2\left(1-F_2\right), \\
        & \frac{dF_2}{dt} = \tilde{\alpha}_1 F_1 F_2 + \tilde{\Gamma}_2^1\,F_1\left(1-F_1\right) + \Gamma_2^2\,F_2\left(1-F_2\right).
    \end{aligned}    
\end{equation}
Therefore, the assumption of unitary carrying capacity in \eqref{sys2} for all involved species is not a restriction on the structure of the model.
\end{remark}

\section{The coexistence point: stability analysis}\label{seccoex}

In this Section, we present some qualitative results, mainly focusing on the linear stability of the coexistence point.

Without loss of generality, we assume that $\Gamma_1^1<0$ (equivalently, one may take $\Gamma_2^2<0$). Then, the kinetic system~\eqref{sys2} admits these equilibria in $\mathbb{R}^2$:
\begin{equation}\label{equilibrium}
E_1=(f_1^\star,f_2^\star), \qquad E_2=(0,0), \qquad E_3=(0,1), \qquad E_4=(1,0).
\end{equation}
As underlined in Section~\ref{sececolg}, we are interested only in strictly positive values of the solution, so that we study only the coexistence equilibrium $E_1$, where
\begin{equation}\label{equilibrium_value}
\begin{aligned}
& f_1^\star=\frac{(\Gamma_1^2 \Gamma_2^1-\Gamma_1^1 \Gamma_2^2)(\Gamma_1^2 \Gamma_2^1-\Gamma_1^1 \Gamma_2^2+\alpha(\Gamma_1^2+\Gamma_2^2) )}{\alpha^2 (\Gamma_1^1+\Gamma_2^2)(\Gamma_1^2+\Gamma_2^2)+(\Gamma_1^2 \Gamma_2^1-\Gamma_1^1 \Gamma_2^2)^2}, \\
& f_2^\star=-\frac{(\Gamma_1^2 \Gamma_2^1-\Gamma_1^1 \Gamma_2^2)(-\Gamma_1^2 \Gamma_2^1+\Gamma_1^1 \Gamma_2^2+\alpha(\Gamma_1^1+\Gamma_2^1) )}{\alpha^2 (\Gamma_1^1+\Gamma_2^2)(\Gamma_1^2+\Gamma_2^2)+(\Gamma_1^2 \Gamma_2^1-\Gamma_1^1 \Gamma_2^2)^2},
\end{aligned}
\end{equation}
fully determined by the parameters of the system~\eqref{sys2}. 

In the following Subsections, we will analyze four different scenarios, by suitably choosing 
the real value $\alpha$ and the components $\Gamma_i^h$ of the external force field involved in the system~\eqref{sys2}. Of course, the different scenarios affect the values of the coexistence equilibrium.

Hereafter, we assume that coefficients $\Gamma_{i}^j$'s have the form
\begin{equation}\label{assump2}
\Gamma_1^2 = a \gamma, \quad \Gamma_2^1 = b \gamma, \quad \Gamma_1^1=c\gamma, \quad \Gamma_2^2=d \gamma,
\end{equation}
where $a$ $b$, $c$, $d$ are real constants. Then, the components of the coexistence equilibrium simplify to
\begin{equation}\label{coexcistence}
\begin{aligned}
 & f_1^\star = \frac{\gamma (ab -cd) \left( \gamma(ab-cd) + \alpha (a+b)\right])}{\gamma^2 (ab-cd)^2 + \alpha^2(a+d)(b+c)}, \\
& f_2^\star = \frac{\gamma (ab -cd) \left( \gamma(ab-cd) - \alpha (b+c)\right) }{\gamma^2 (ab-cd)^2 + \alpha^2(a+d)(b+c)}.
\end{aligned}
\end{equation}
Moreover, in the first three scenarios, \emph{i.e.}, those ones analyzed in Subsections~\ref{subsecfirts}--\ref{subsecthi}, we set
\begin{equation}\label{assump3}
\alpha=g \gamma,    
\end{equation}
whereupon, the components of the coexistence equilibrium $E_1$ reduce to
\begin{equation}
\label{coexcistence_4casi}
\begin{aligned}
& f_1^\star = \frac{(ab -cd) \left[ (ab-cd) + g (a+d) \right]}{(ab-cd)^2 + g^2(a+d)(b+c)}, \\
& f_2^\star = \frac{(ab -cd) \left[ (ab-cd) - g (b+c) \right]}{(ab-cd)^2 + g^2 (a+d)(b+c)}.
\end{aligned}
\end{equation}
The linear stability is obtained by analyzing the eigenvalues of the Jacobian matrix
\begin{equation}
\label{jacobian_matrix}
    J|_{(f_1^\star,f_2^\star)} =
    \left(
        \begin{array}{cc}
             -\alpha f_2^\star + (1-2\,f_1^\star) \Gamma_1^1 & -\alpha f_1^\star + (1-2\,f_2^\star) \Gamma_1^2 \\
             \alpha f_2^\star + (1-2\,f_1^\star) \Gamma_2^1 & \alpha f_1^\star + (1-2\,f_2^\star) \Gamma_2^2 \\
        \end{array}
    \right)
\end{equation}
Finally, the Matlab routine \texttt{ode45} has been used for the numerical integration.

\subsection{First scenario}\label{subsecfirts}
\begin{figure}
    \centering
    \subfigure[]{\includegraphics[width=0.45\textwidth]{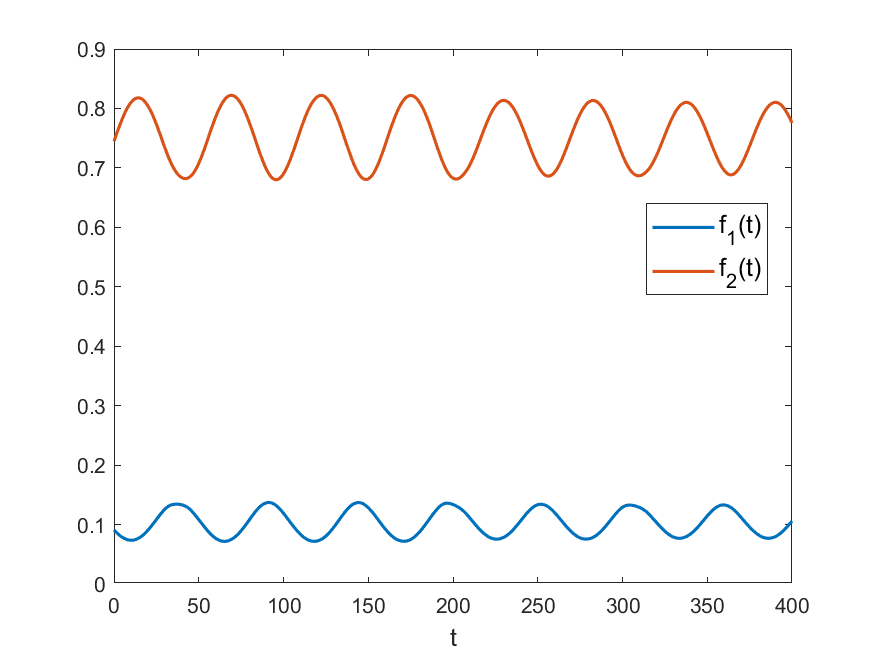}}
    \subfigure[]{\includegraphics[width=0.45\textwidth]{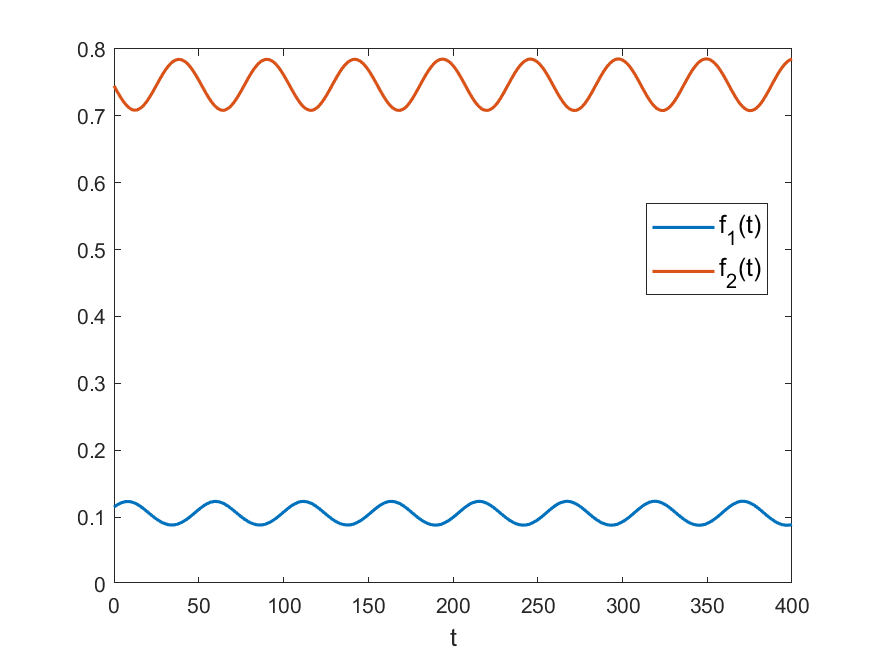}} \\
    \subfigure[]{\includegraphics[width=0.45\textwidth]{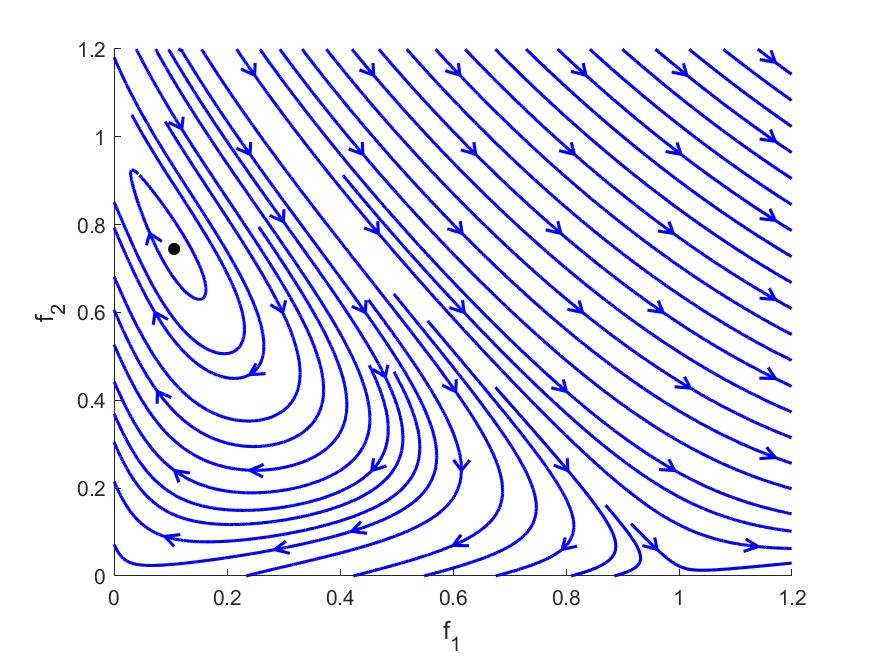}}
    \subfigure[]{\includegraphics[width=0.45\textwidth]{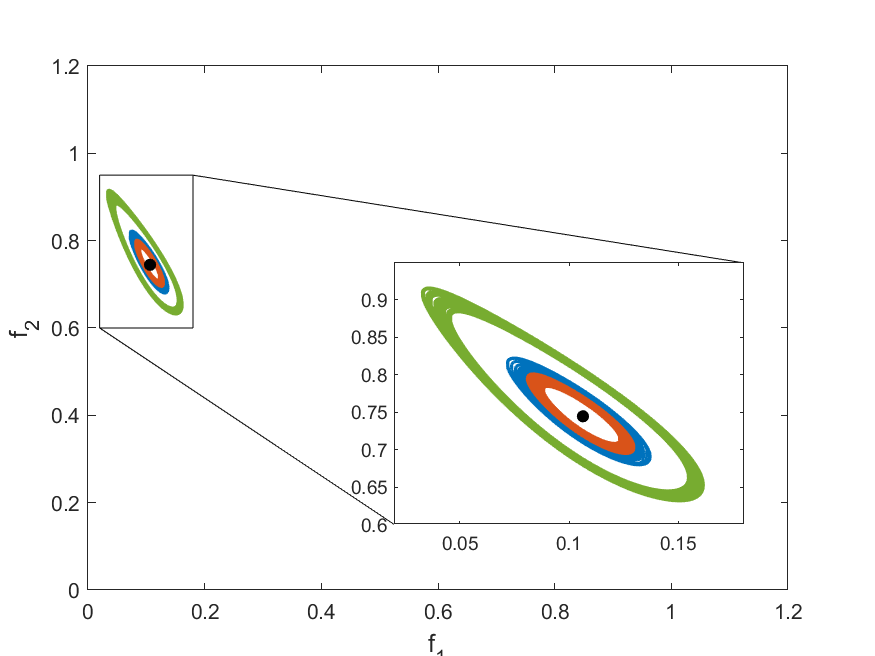}} 
    \caption{\label{first_scenario}Time evolution of $f_1(t)$ and $f_2(t)$ in the time interval $[0,400]$, with different initial conditions (a) $\mathbf{f^0}=(0.0910,0.7450)$ and (b) $\mathbf{f^0}=(0.1150,0.7450)$. Subfigure (c) displays the phase portrait, and subfigure (d) the trajectories obtained by solving the system with different initial conditions chosen in the ``basin of attraction" of $E_1$. The black dots represent the equilibria.}
\end{figure}

In this first scenario, the values of the parameters~\eqref{assump2} and \eqref{assump3} are fixed as follows:
\begin{equation}\label{eqparcoex}
a=-1, \quad b=-1, \quad  c=-3, \quad  d=3, \quad g=-6,
\end{equation}
The coexistence equilibrium point specializes to
\begin{equation*}
f_1^\star=0.1064,\qquad f_2^\star=0.7447,
\end{equation*}
and the eigenvalues of the Jacobian matrix~\eqref{jacobian_matrix} are
\begin{equation*}
 \lambda_{\pm}=\pm 1.2204\,\gamma\, \textrm{i} .
\end{equation*}
Then, the coexistence equilibrium point $E_1$ is a center. Choosing
$\gamma=0.1$,
the system~\eqref{sys2} admits, at least locally in time, a positive and bounded solution $\mathbf{f}(t)=(f_1(t),\, f_2(t))$. Some numerical simulations have been provided in this first scenario (see Figure~\ref{first_scenario}), where different initial conditions have been considered. In particular, the existence of periodic solutions is evident from the plots. The basin of attraction, which does not seem to be very large, of the coexistence equilibrium $E_1$ is also represented.

\subsection{Second scenario}\label{subsecsec}
\begin{figure}
    \centering
    \subfigure[]{\includegraphics[width=0.45\textwidth]{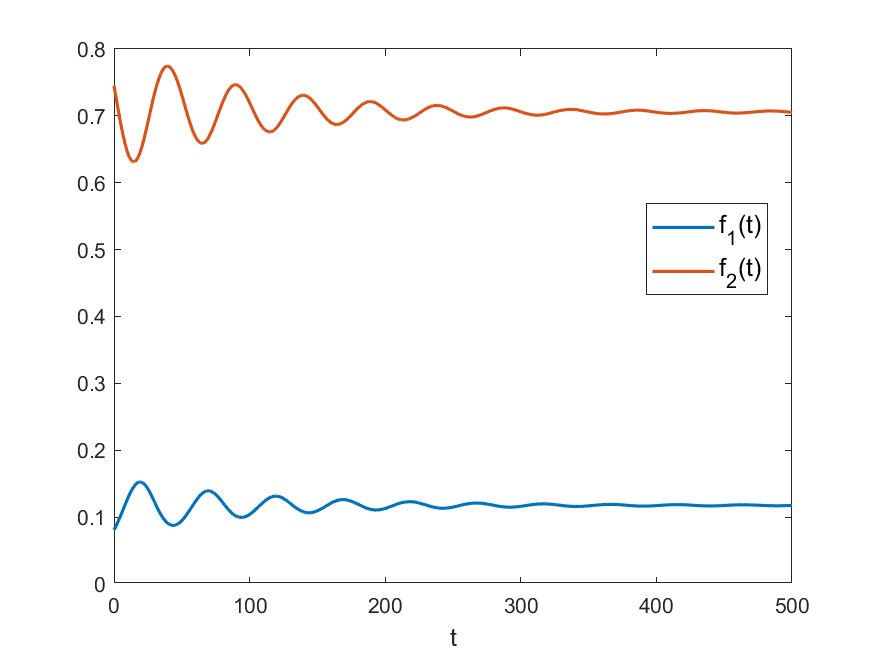}}
    \subfigure[]{\includegraphics[width=0.45\textwidth]{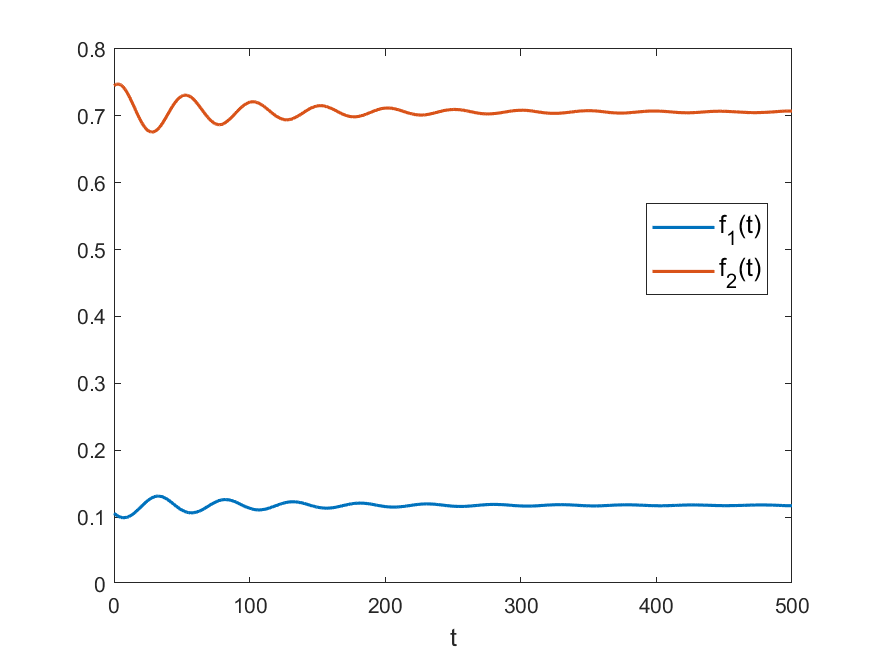}} \\
    \subfigure[]{\includegraphics[width=0.45\textwidth]{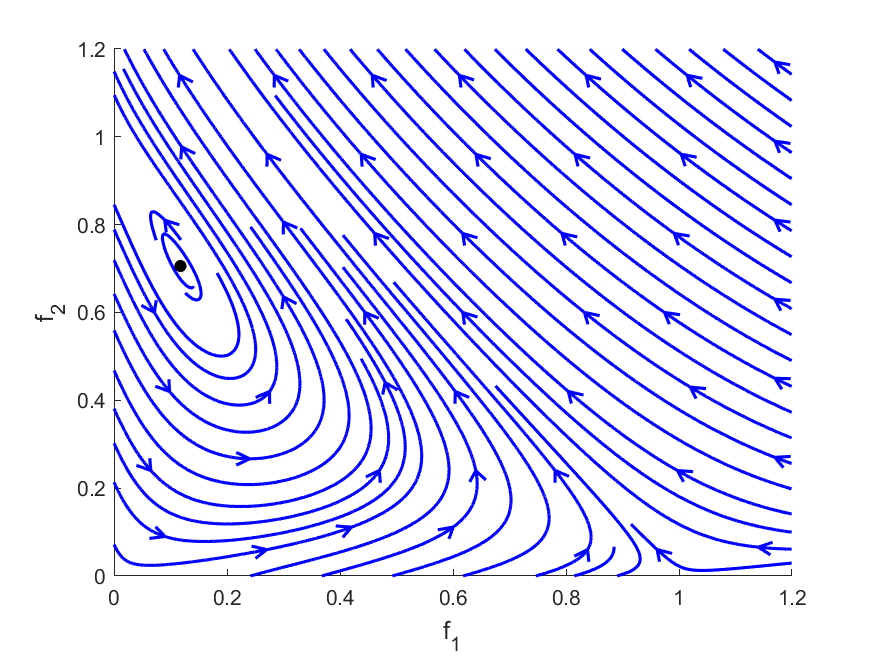}}
    \subfigure[]{\includegraphics[width=0.45\textwidth]{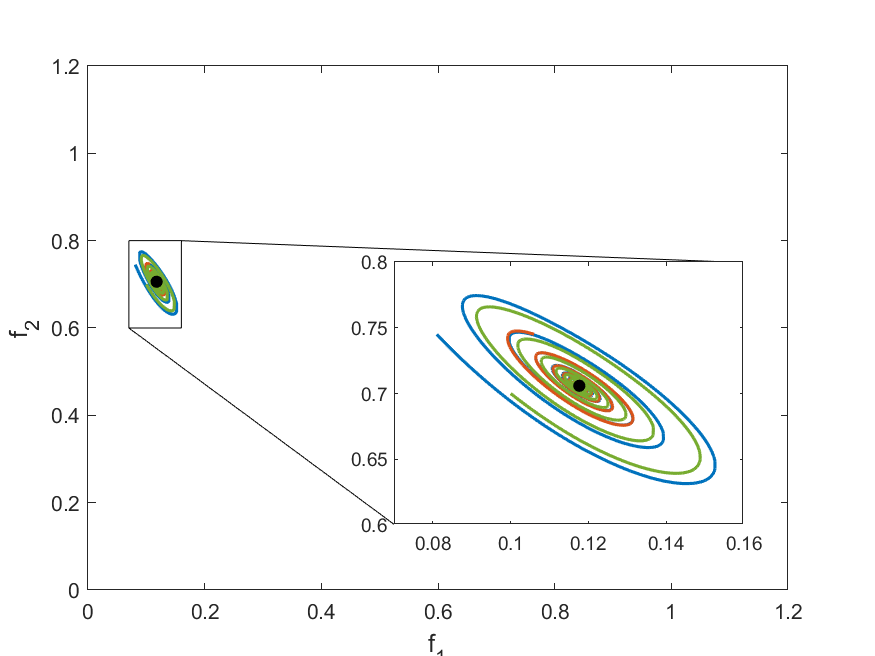}}
\caption{\label{second_scenario}
Time evolution of $f_1(t)$ and $f_2(t)$ in the time interval $[0,500]$, with different initial conditions (a) $\mathbf{f^0}=(0.0810,0.7450)$ and (b) $\mathbf{f^0}=(0.1060,0.7450)$. Subfigure (c) displays the phase portrait, and subfigure (d) the trajectories obtained by solving the system with different initial conditions chosen in the ``basin of attraction" of $E_1$. The black dots represent the equilibria.}
\end{figure}

In the second scenario, the constants involved in the expressions~\eqref{assump2} and \eqref{assump3} are fixed as follows:
\begin{equation*}
    a=-1, \quad b=-1, \quad  c=-2.8, \quad  d=2.9, \quad g=-6.
\end{equation*}
By evaluating the Jacobian matrix~\eqref{jacobian_matrix} in the coexistence equilibrium $E_1$, we obtain the eigenvalues
\begin{equation*}
\lambda_{\pm}=0.0971\,\gamma \pm 1.2649\,\gamma\,\textrm{i} .
\end{equation*}
If $\gamma <0$ the equilibrium $E_1$ is a stable spiral sink, whereas, if $\gamma >0$, $E_1$ is unstable.
Once again, let us choose $\gamma=-0.1$, so that
the system~\eqref{sys2} admits a unique positive solution $\mathbf{f}(t)=(f_1(t),\, f_2(t))$, at least local in time. The equilibrium $E_1=(f_1^\star,\, f_2^\star)$ numerically evaluates to
\begin{equation*}
f_1^\star=0.1176,\qquad
f_2^\star=0.7059.
\end{equation*}
The results of the numerical simulations are displayed in Figure~\ref{second_scenario}, where different initial conditions are considered. Of course, the evolution exhibits damped oscillations. Moreover, the basin of attraction is shown, and also in this scenario, it is not very large.

\subsection{Third scenario}\label{subsecthi}
\begin{figure}
 \centering
 \subfigure[]{\includegraphics[width=0.45\textwidth]{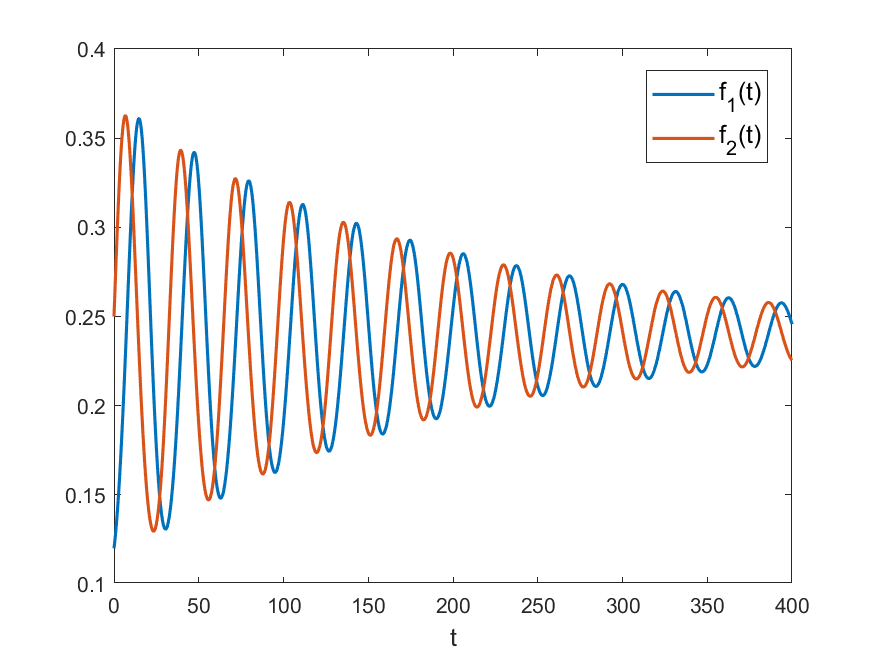}}
 \subfigure[]{\includegraphics[width=0.45\textwidth]{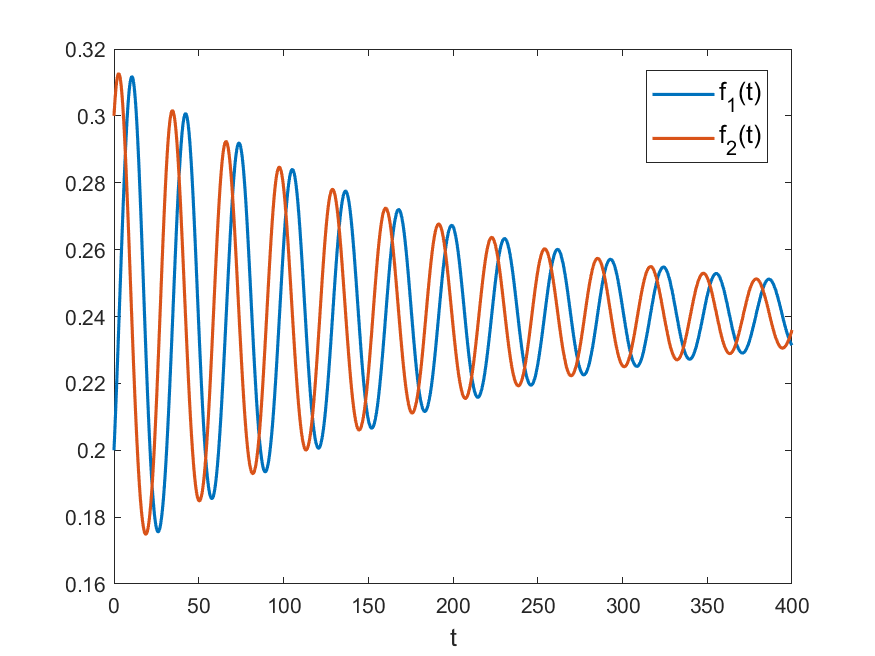}} \\
 \subfigure[]{\includegraphics[width=0.45\textwidth]{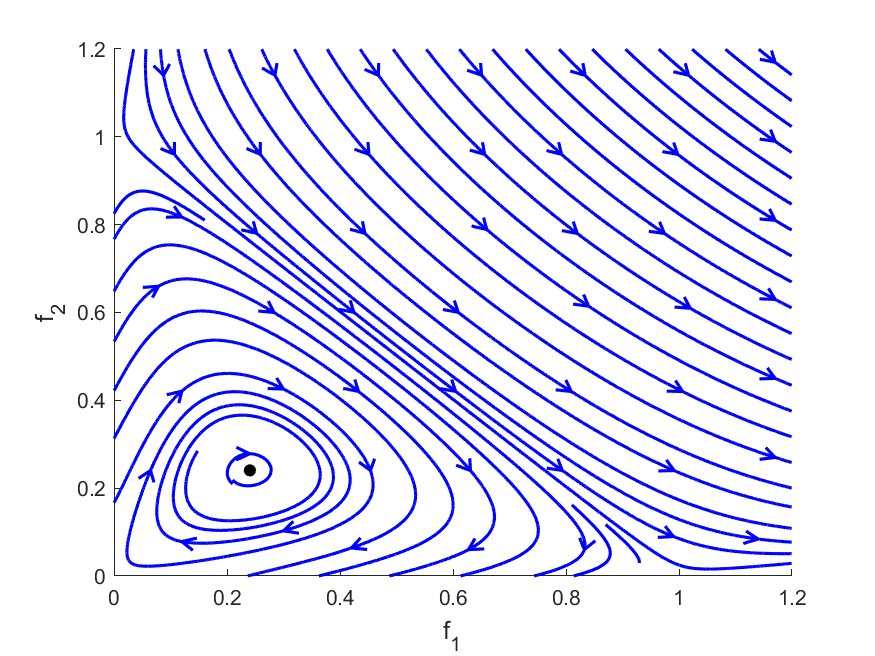}}
 \subfigure[]{\includegraphics[width=0.45\textwidth]{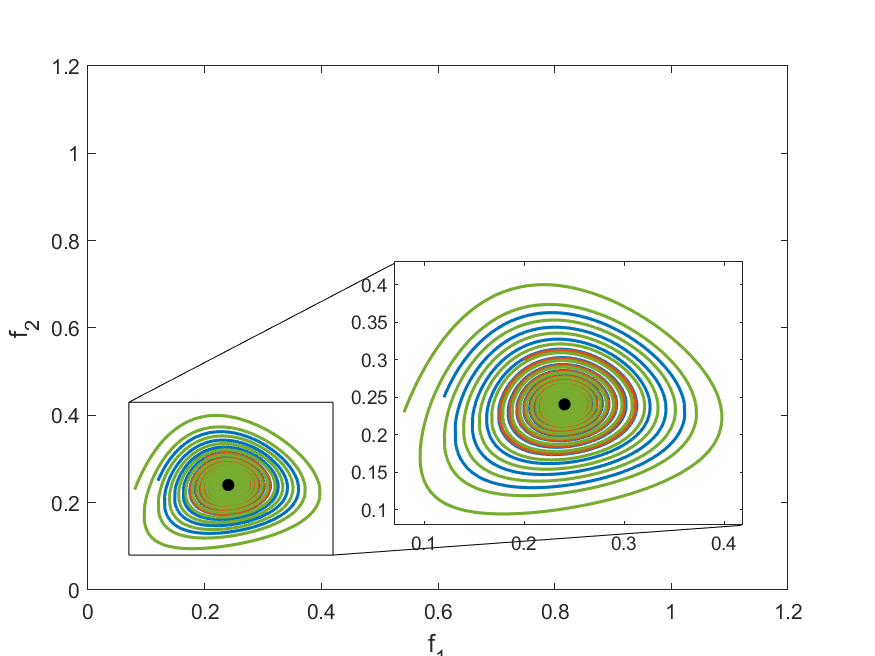}} 
 \caption{\label{third_scenario}Time evolution of $f_1(t)$ and $f_2(t)$ in the time interval $[0,400]$, with different initial conditions (a) $\mathbf{f^0}=(0.1200,0.2500)$ and (b) $\mathbf{f^0}=(0.2000,0.3000)$. Subfigure (c) displays the phase portrait, and subfigure (d) the trajectories obtained by solving the system with different initial conditions chosen in the ``basin of attraction" of $E_1$. The black dots represent the equilibria.}
\end{figure}

In the third scenario, the parameters involved in \eqref{assump2} and \eqref{assump3} are assumed to be 
\begin{equation*}
    a=1.2, \quad b=-1, \quad  c=-3.1, \quad  d=2.9, \quad g=-6.
\end{equation*}
Then, the eigenvalues of the Jacobian matrix~\eqref{jacobian_matrix} are
\begin{equation*}
\lambda_{\pm}=-0.0519\,\gamma \pm 2.0100\,\gamma\,\textrm{i} .
\end{equation*}
Specifically: if $\gamma>0$ the equilibrium $E_1$ is stable, whereas, if $\gamma<0$, the equilibrium $E_1$ is unstable. 
Along the choice $\gamma=0.1$,
the system~\eqref{sys2} admits a unique positive solution $\mathbf{f}(t)=(f_1(t),\, f_2(t))$, at least locally in time. Moreover, the equilibrium $E_1$ specializes to
\begin{equation*}
 f_1^\star=0.2405,\qquad
 f_2^\star=0.2405.
\end{equation*}
Figure~\ref{third_scenario} displays some numerical simulations; as expected, damped oscillations are exhibited, and they persist for longer times. Moreover, the evolution of the two populations shows values of the densities very close each other. Remarkably,  in this case the basin of attraction of the coexistence equilibrium $E_1$ is wider. 

\subsection{Fourth scenario}\label{subsecfou}
This fourth scenario is qualitatively different with respect to the previous ones. First, the values of the parameters in \eqref{assump2} are assumed to be
\begin{equation}\label{assump5}
 a=-2, \quad b=1, \quad c=1, \quad d=-1,\quad \gamma=-0.1.
 \end{equation}
The coexistence equilibrium point specializes to
\begin{equation}\label{coex4}
 f_1^\star=\frac{1-30\alpha}{1-600\alpha^2},\qquad
 f_2^\star=\frac{1-20\alpha}{1-600\alpha^2},
\end{equation}
which is meaningful provided that $-1/(10\sqrt{6})<\alpha<1/30$ or $\alpha>1/20$,
and its nature is described by Theorem~\ref{teo_fourth_scenario}.
\begin{figure}
 \centering
 \subfigure[]{\includegraphics[width=0.45\textwidth]{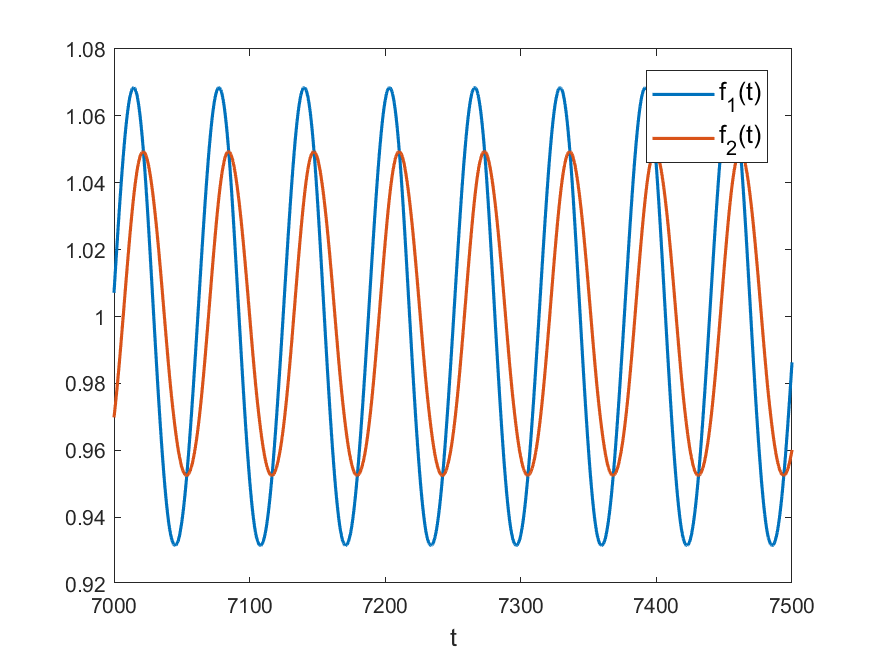}}
 \subfigure[]{\includegraphics[width=0.45\textwidth]{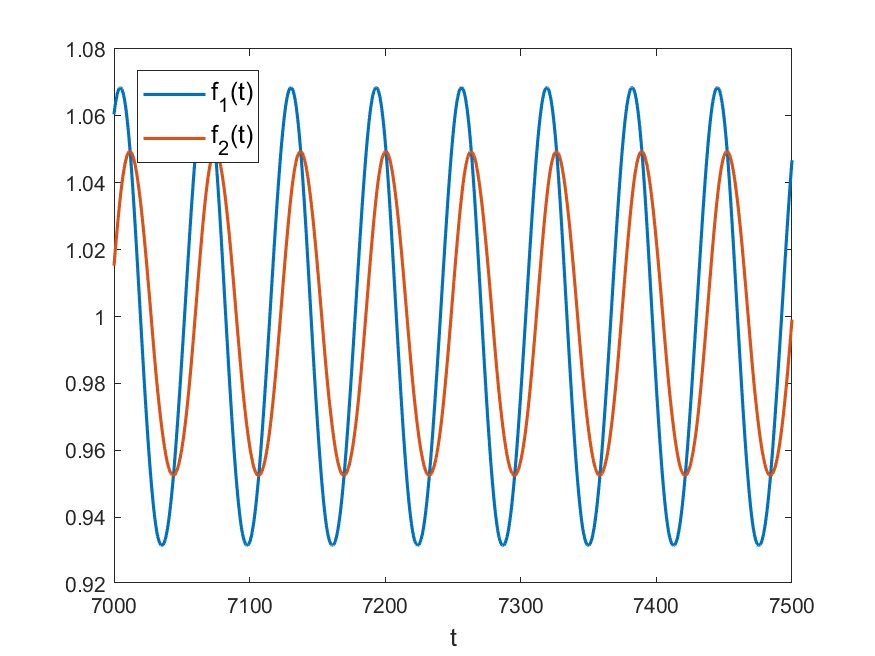}} \\
 \subfigure[]{\includegraphics[width=0.45\textwidth]{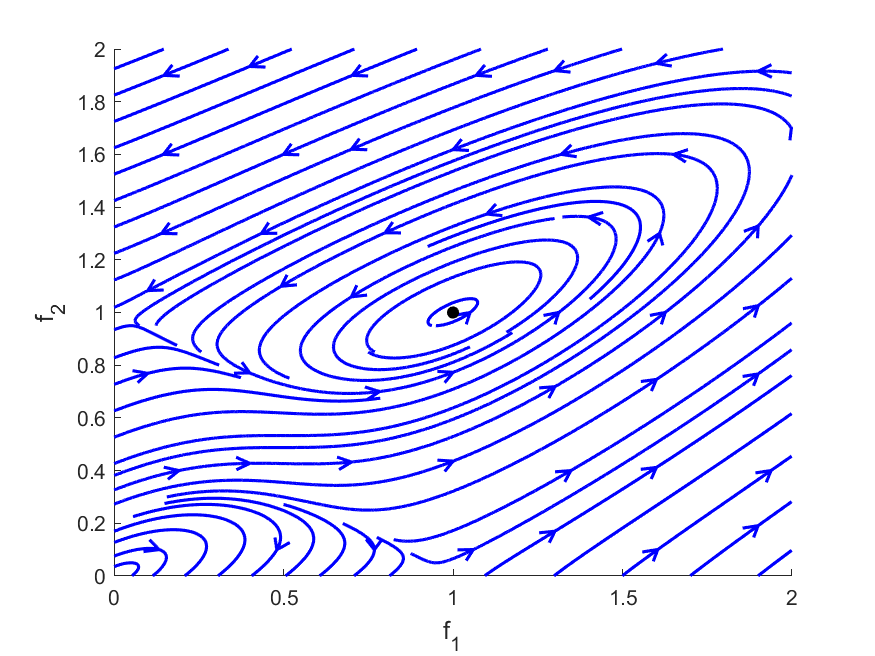}}
 \subfigure[]{\includegraphics[width=0.45\textwidth]{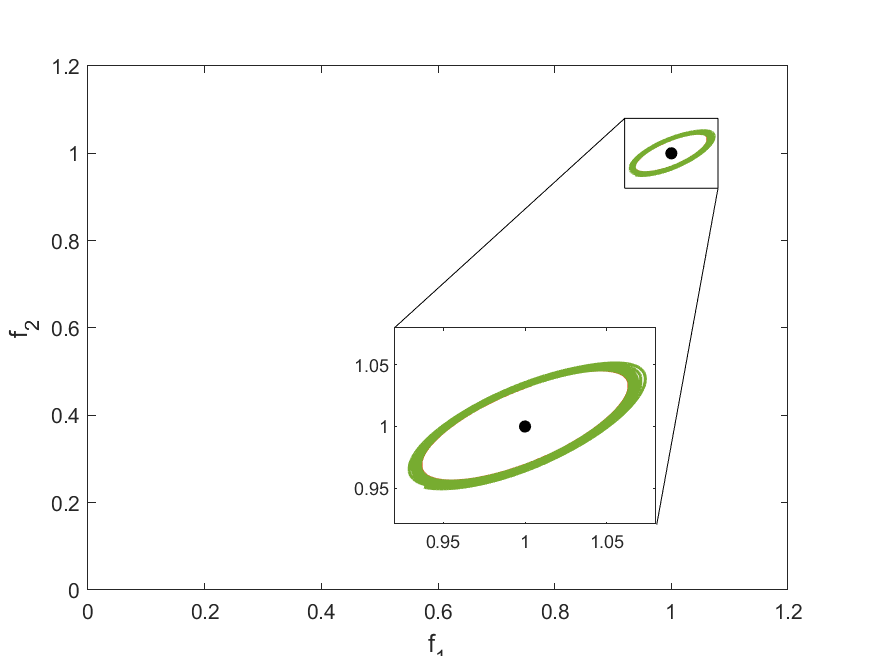}} \\
 \subfigure[]{\includegraphics[width=0.45\textwidth]{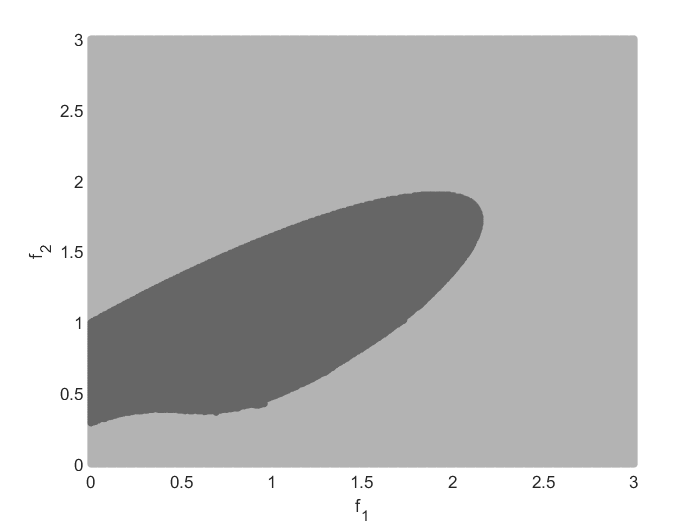}}
 \caption{\label{fourth_scenario}Time evolution of $f_1(t)$ and $f_2(t)$ in the time interval $[7000,7500]$, with different initial conditions (a) $\mathbf{f^0}=(0.9,0.9)$ and (b) $\mathbf{f^0}=(0.5,0.8)$. Subfigure (c) displays the phase portrait, and subfigure (d) the trajectories obtained by solving the system with different initial conditions chosen in the ``basin of attraction" of $E_1$
(shown in subfigure (e)). The black dots represent the equilibria. It is possible to see that around the equilibria there are stable limit cycles.}
\end{figure}

\begin{theorem}{(Hopf bifurcation).}\label{teo_fourth_scenario}
The equilibrium point $E_1$, given in \eqref{coex4}, provided that $-1/(10\sqrt{6})<\alpha<1/30$ or $\alpha>1/20$, undergoes, for $\alpha=0$, a Hopf bifurcation.
\begin{proof}
The eigenvalues of \eqref{jacobian_matrix} are 
\begin{equation*} 
\lambda_{\pm}=\frac{10\alpha(1+5\alpha) \pm \sqrt{-1+50\alpha(1+2\alpha(1+145\alpha(-2+25\alpha)))}}{6000\alpha^2-10}.
\end{equation*}
It is easy to verify that we have complex conjugate eigenvalues:  when $-0.0402916 < \alpha < 0.0249581$. Moreover, for $0<\alpha<0.0249581$, the eigenvalues have negative real part $E_1$ stable spiral point), whereas for $-0.0402916 < \alpha < 0$ have positive real part ($E_1$ unstable spiral point); finally, for $\alpha=0$ we have purely imaginary eigenvalues. The derivative of the real part of these eigenvalues, evaluated at the bifurcation value $\alpha=0$, is not zero, \emph{i.e.}, 
\begin{equation}\label{hpf}
\frac{d Re(\lambda_{\pm})}{d \alpha}\bigg|_ {\alpha=0} = -\frac{ \gamma^2 (\gamma^2 - \alpha \gamma + 6\alpha^2)}{(\gamma^2-6\alpha^2)^2}\bigg|_ {\alpha=0} = -1\neq 0.
 \end{equation}
 Thus,  \eqref{hpf} implies that at $\alpha=0$ we have a Hopf bifurcation.
 \end{proof}
\end{theorem}

Notice that at $\alpha=0$, where the Hopf bifurcation occurs, the coexistence equilibrium becomes
$E_1=(1, 1)$.

In Figure~\ref{fourth_scenario}, the numerical simulations concerned with this fourth scenario are provided. In particular, the basin of attraction of the coexistence equilibrium is shown, and periodic patterns appear according to the Hopf bifurcation described in Theorem~\ref{teo_fourth_scenario}. Moreover, in Figure~\ref{fourth_scenario}, we show both the periodic orbits and the basin of attraction of the coexistence equilibrium.

\section{Conclusions}\label{SecConcl}
In this paper, we have developed a nonconservative kinetic system that mimics the evolution of an ecological system composed of $n$ population. After defining the specific analytical form of the external force field~\eqref{shapeforc}, the kinetic equations~\eqref{eqnewfor} have been derived. In particular, the external force field includes the effect of the logistic growth of each population on all the others. At the best of our knowledge, this represents a first attempt for deriving such ecological framework by using a kinetic system subjected to an external force field. Specifically, in this case, the external force field is the natural logistic growth evolution of species, \emph{i.e.}, the system is seen as \emph{open}, and not only as a \emph{closed} system subjected to binary and stochastic interactions.

After deriving the kinetic system~\eqref{eqnewfor}, we gave its ecological interpretation. In order to carry out numerical simulations, we have focused our attention on a system consisting of only two populations, each with its logistic growth and the related impact on both populations. Although this could seem restrictive, this allowed us to better identify certain features of the model, along with its descriptive potential. Therefore, we have obtained the final kinetic system, that is a system of nonlinear ordinary differential equations. We have discussed the analytical properties of equilibria. In particular, we have considered the coexistence equilibrium and its linear stability properties. The interest towards the coexistence equilibrium has not only mathematical reasons; indeed, the coexistence represents an important aspect for ecological reasons. Different patterns are exhibited. In one scenario the coexistence point is a center, whereas in the other scenarios is a spiral. Finally, we considered a fourth scenario where a Hopf bifurcation may be detected. 

We remark that this paper represents a first kinetic modeling approach to ecological dynamics, by taking into account still the growth of the involved populations. It is worth stressing that the main novelty is the modeling approach through an external force field of a logistic growth of the interacting species. This can represent an extension of classical Lotka-Volterra models, widely used in the last years.  As a future research perspective, there are at least two further aspects to investigate. On one hand, a spatial inhomogeneous model can be introduced in order to have a more accurate description of the phenomenon. Indeed, the evolution of a population strongly depends on its geographic location, along with the respective movements over time (see \cite{austin2002spatial,chen2011spatio,hastings1990spatial}, and references therein). From a mathematical viewpoint, this choice means that the related kinetic system will be composed of partial-differential equations or integro-partial-differential equations, unlike what happens with system~\eqref{eqnewfor}. This could imply some further mathematical difficulties, concerned first with the well-posedness of the problem. Maybe, the introduction of a spatial variable could lead to the emergence of Turing instabilities. On the other hand, we aim to investigate the cases where one of the involved populations may go to extinction, as this occurrence is neglected in this work. The problem of extinction has a great interest \cite{sudakow2022knowledge}, also with respect to climate change emergence \cite{sekerci2020climate}. Results on both directions are in progress. Finally, the study of global properties of the model~\eqref{eqnewfor} here derived could be of interest, for instance the investigation of analytical conditions that ensure global existence and uniqueness of solutions even if the overall evolution proceeds in a nonconservative scheme.

\medskip
\paragraph*{Acknowledgements.} 
Work supported by G.N.F.M. of I.N.d.A.M.

\bibliographystyle{unsrt}
\bibliography{biblio}

@article{tosin2019kinetic,
  title={Kinetic-controlled hydrodynamics for traffic models with driver-assist vehicles},
  author={Tosin, A. and Zanella, M.},
  journal={Multiscale Modeling \& Simulation},
  volume={17},
  number={2},
  pages={716--749},
  year={2019},
  publisher={SIAM}
}

@article{bertotti2004discrete,
  title={From discrete kinetic and stochastic game theory to modelling complex systems in applied sciences},
  author={Bertotti, M.~L. and Delitala, M.},
  journal={Mathematical Models and Methods in Applied Sciences},
  volume={14},
  number={07},
  pages={1061--1084},
  year={2004},
  publisher={World Scientific}
}

@article{arlotti1999jager,
  title={From the Jager and Segel model to kinetic population dynamics nonlinear evolution problems and applications},
  author={Arlotti, L. and Bellomo, N. and Latrach, K.},
  journal={Mathematical and Computer Modelling},
  volume={30},
  number={1-2},
  pages={15--40},
  year={1999},
  publisher={Elsevier}
}

@article{carbonaro2023nonconservative,
  title={A nonconservative kinetic framework under the action of an external force field: Theoretical results with application inspired to ecology},
  author={Carbonaro, B. and Menale, M.},
  journal={European Journal of Applied Mathematics},
  volume={34},
  number={6},
  pages={1170--1186},
  year={2023},
  publisher={Cambridge University Press}
}

@article{menale2023kinetic,
  title={A kinetic framework under the action of an external force field: Analysis and application in epidemiology},
  author={Menale, M. and Munaf{\`o}, C.~ F.},
  journal={Chaos, Solitons \& Fractals},
  volume={174},
  pages={113801},
  year={2023},
  publisher={Elsevier}
}

@book{coddington2012introduction,
  title={An introduction to ordinary differential equations},
  author={Coddington, E.~A.},
  year={2012},
  publisher={Courier Corporation}
}

@article{pareschi2019hydrodynamic,
  title={Hydrodynamic models of preference formation in multi-agent societies},
  author={Pareschi, L. and Toscani, G. and Tosin, A. and Zanella, M.},
  journal={Journal of Nonlinear Science},
  volume={29},
  pages={2761--2796},
  year={2019},
  publisher={Springer}
}

@article{toscani2006kinetic,
  title={Kinetic models of opinion formation},
  author={Toscani, G.},
  journal={Communications in Mathematical Sciences},
  volume={4},
  number={3},
  pages={481--496},
  year={2006},
  publisher={International Press of Boston}
}

@article{toscani2018opinion,
  title={Opinion modeling on social media and marketing aspects},
  author={Toscani, G. and Tosin, A. and Zanella, M.},
  journal={Physical Review E},
  volume={98},
  number={2},
  pages={022315},
  year={2018},
  publisher={APS}
}

@article{bellomo2008complexity,
  title={On the complexity of multiple interactions with additional reasonings about Kate, Jules and Jim},
  author={Bellomo, N. and Carbonaro, B.},
  journal={Mathematical and Computer Modelling},
  volume={47},
  number={1-2},
  pages={168--177},
  year={2008},
  publisher={Elsevier}
}

@article{albi2019vehicular,
  title={Vehicular traffic, crowds, and swarms: From kinetic theory and multiscale methods to applications and research perspectives},
  author={Albi, G. and Bellomo, N. and Fermo, L. and Ha, S-Y and Kim, J. and Pareschi, L. and Poyato, D. and Soler, J.},
  journal={Mathematical Models and Methods in Applied Sciences},
  volume={29},
  number={10},
  pages={1901--2005},
  year={2019},
  publisher={World Scientific}
}

@article{puppo2017analysis,
  title={Analysis of a multi-population kinetic model for traffic flow},
  author={Puppo, G. and Semplice, M. and Tosin, A. and Visconti, G.},
  journal={Communications in Mathematical Sciences},
  volume={15},
  number={2},
  pages={379--412},
  year={2017},
  publisher={International Press of Boston}
}

@article{della2023sir,
  title={An SIR model with viral load-dependent transmission},
  author={Della Marca, R. and Loy, N. and Tosin, A.},
  journal={Journal of Mathematical Biology},
  volume={86},
  number={4},
  pages={61},
  year={2023},
  publisher={Springer}
}

@article{loy2021viral,
  title={A viral load-based model for epidemic spread on spatial networks},
  author={Loy, N. and Tosin, A.},
  journal={Mathematical Biosciences and Engineering},
  volume={18},
  number={5},
  pages={5635--5663},
  year={2021}
}

@incollection{albi2022kinetic,
  title={Kinetic modelling of epidemic dynamics: social contacts, control with uncertain data, and multiscale spatial dynamics},
  author={Albi, G. and Bertaglia, G. and Boscheri, W. and Dimarco, G. and Pareschi, L. and Toscani, G. and Zanella, M.},
  booktitle={Predicting Pandemics in a Globally Connected World, Volume 1: Toward a Multiscale, Multidisciplinary Framework through Modeling and Simulation},
  pages={43--108},
  year={2022},
  publisher={Springer}
}

@article{bernardi2022effects,
  title={Effects of vaccination efficacy on wealth distribution in kinetic epidemic models},
  author={Bernardi, E. and Pareschi, L. and Toscani, G. and Zanella, M.},
  journal={Entropy},
  volume={24},
  number={2},
  pages={216},
  year={2022},
  publisher={MDPI}
}

@article{dimarco2020wealth,
  title={Wealth distribution under the spread of infectious diseases},
  author={Dimarco, G. and Pareschi, L. and Toscani, G. and Zanella, M.},
  journal={Physical Review E},
  volume={102},
  number={2},
  pages={022303},
  year={2020},
  publisher={APS}
}

@article{della2023intransigent,
  title={Intransigent vs. volatile opinions in a kinetic epidemic model with imitation game dynamics},
  author={Della Marca, R. and Loy, N. and Menale, M.},
  journal={Mathematical Medicine and Biology: A Journal of the IMA},
  volume={40},
  number={2},
  pages={111--140},
  year={2023},
  publisher={Oxford University Press}
}

@article{bellomo2014multiscale,
  title={On the multiscale modeling of vehicular traffic: from kinetic to hydrodynamics},
  author={Bellomo, N. and Bellouquid, A. and Nieto, J. and Solerz, J.},
  journal={Discrete \& Continuous Dynamical Systems-Series B},
  volume={19},
  number={7},
  year={2014}
}

@book{bellomo2008modeling,
  title={Modeling complex living systems: a kinetic theory and stochastic game approach},
  author={Bellomo, N.},
  year={2008},
  publisher={Springer Science \& Business Media}
}

@article{bertotti2012exploiting,
  title={Exploiting the flexibility of a family of models for taxation and redistribution},
  author={Bertotti, M.~L. and Modanese, G.},
  journal={The European Physical Journal B},
  volume={85},
  pages={1--10},
  year={2012},
  publisher={Springer}
}

@article{letizia2016economic,
  title={Economic inequality and mobility in kinetic models for social sciences},
  author={Bertotti, M.~L. and Modanese, G.},
  journal={The European Physical Journal Special Topics},
  volume={225},
  pages={1945--1958},
  year={2016},
  publisher={Springer}
}

@article{bertotti2017stochastic,
  title={Stochastic effects in a discretized kinetic model of economic exchange},
  author={Bertotti, M.~L. and Chattopadhyay, A.~K. and Modanese, G.},
  journal={Physica A: Statistical Mechanics and its Applications},
  volume={471},
  pages={724--732},
  year={2017},
  publisher={Elsevier}
}

@article{bellomo2009complexity,
  title={Complexity analysis and mathematical tools towards the modelling of living systems},
  author={Bellomo, N. and Bianca, C. and Delitala, M.},
  journal={Physics of Life Reviews},
  volume={6},
  number={3},
  pages={144--175},
  year={2009},
  publisher={Elsevier}
}

@article{bertotti2008discrete,
  title={On a discrete generalized kinetic approach for modelling persuader’s influence in opinion formation processes},
  author={Bertotti, M.~L. and Delitala, M.},
  journal={Mathematical and Computer Modelling},
  volume={48},
  number={7-8},
  pages={1107--1121},
  year={2008},
  publisher={Elsevier}
}

@article{tsoularis2002analysis,
  title={Analysis of logistic growth models},
  author={Tsoularis, A. and Wallace, J.},
  journal={Mathematical Biosciences},
  volume={179},
  number={1},
  pages={21--55},
  year={2002},
  publisher={Elsevier}
}

@article{mandal2020nonautonomous,
  title={A nonautonomous model for the effect of environmental toxins on plankton dynamics},
  author={Mandal, A. and Tiwari, P.~K. and Samanta, S. and Venturino, E. and Pal, S.},
  journal={Nonlinear Dynamics},
  volume={99},
  pages={3373--3405},
  year={2020},
  publisher={Springer}
}

@article{menale2024kinetic,
  title={A kinetic theory approach to modeling prey--predator ecosystems with expertise levels: analysis, simulations and stability considerations},
  author={Menale, M. and Venturino, E.},
  journal={Computational and Applied Mathematics},
  volume={43},
  number={4},
  pages={216},
  year={2024},
  publisher={Springer}
}

@article{toscani2023kinetic,
  title={On a kinetic description of {L}otka-{V}olterra dynamics},
  author={Toscani, G. and Zanella, M.},
  journal={arXiv preprint arXiv:2302.14573},
  year={2023}
}

@article{lotka2002contribution,
  title={Contribution to the theory of periodic reactions},
  author={Lotka, A.~J.},
  journal={The Journal of Physical Chemistry},
  volume={14},
  number={3},
  pages={271--274},
  year={2002},
  publisher={ACS Publications}
}

@book{volterra1926variazioni,
  title={Variazioni e fluttuazioni del numero d'individui in specie animali conviventi, memoria del socio Vito Volterra},
  author={Volterra, V.},
  year={1926},
  publisher={Societ{\`a} anonima tipografica Leonardo da Vinci}
}

@article{volterra1928variations,
  title={Variations and fluctuations of the number of individuals in animal species living together},
  author={Volterra, V.},
  journal={ICES Journal of Marine Science},
  volume={3},
  number={1},
  pages={3--51},
  year={1928},
  publisher={Oxford University Press}
}

@article{chattopadhyay2008patchy,
  title={Patchy agglomeration as a transition from monospecies to recurrent plankton blooms},
  author={Chattopadhyay, J. and Chatterjee, S. and Venturino, E.},
  journal={Journal of Theoretical Biology},
  volume={253},
  number={2},
  pages={289--295},
  year={2008},
  publisher={Elsevier}
}

@article{bellouquid2011kinetic,
	title={From kinetic models of multicellular growing systems to macroscopic biological tissue models},
	author={Bellouquid, A. and De Angelis, E.},
	journal={Nonlinear Analysis: Real World Applications},
	volume={12},
	number={2},
	pages={1111--1122},
	year={2011},
	publisher={Elsevier}
}

@article{bianca2017modeling,
	title={Modeling the antigen recognition by B-cell and T-cell receptors through thermostatted kinetic theory methods},
	author={Bianca, C. and Br{\'e}zin, L.},
	journal={International Journal of Biomathematics},
	volume={10},
	number={05},
	pages={1750072},
	year={2017},
	publisher={World Scientific}
}

@article{bianca2024decade,
	title={A decade of thermostatted kinetic theory models for complex active matter living systems},
	author={Bianca, C.},
	journal={Physics of Life Reviews},
	year={2024},
	publisher={Elsevier}
}

@article{bianca2024discrete,
	title={In the discrete thermostatted kinetic theory: Proliferations, mutations and microscopic agents},
	author={Bianca, C.},
	journal={Nonlinear Studies},
	volume={31},
	number={3},
	pages={827--837},
	year={2024}
}

@article{ajraldi2011modeling,
	title={Modeling herd behavior in population systems},
	author={Ajraldi, V. and Pittavino, M. and Venturino, E.},
	journal={Nonlinear Analysis: Real World Applications},
	volume={12},
	number={4},
	pages={2319--2338},
	year={2011},
	publisher={Elsevier}
}

@article{chen2023dynamic,
	title={Dynamic complexity of a modified Leslie--Gower predator--prey system with fear effect},
	author={Chen, M. and Takeuchi, Y. and Zhang, J.-F.},
	journal={Communications in Nonlinear Science and Numerical Simulation},
	volume={119},
	pages={107109},
	year={2023},
	publisher={Elsevier}
}

@article{chowdhury2022canards,
	title={Canards, relaxation oscillations, and pattern formation in a slow-fast ratio-dependent predator-prey system},
	author={Chowdhury, P. R. and Banerjee, M. and Petrovskii, S.},
	journal={Applied Mathematical Modelling},
	volume={109},
	pages={519--535},
	year={2022},
	publisher={Elsevier}
}

@book{malchow2007spatiotemporal,
	title={Spatiotemporal patterns in ecology and epidemiology: theory, models, and simulation},
	author={Malchow, H. and Petrovskii, S.~V. and Venturino, E.},
	year={2007},
	publisher={Chapman and Hall/CRC}
}

@article{laurie2020herding,
	title={Herding induced by encounter rate, with predator pressure influencing prey response},
	author={Laurie, H. and Venturino, E. and Bulai, I.~ M.},
	journal={Current Trends in Dynamical Systems in Biology and Natural Sciences},
	pages={63--93},
	year={2020},
	publisher={Springer}
}

@article{liang2022nonlocal,
	title={Nonlocal interactions between vegetation induce spatial patterning},
	author={Liang, J. and Liu, C. and Sun, G.-Q. and Li, L. and Zhang, L. and Hou, M. and Wang, H. and Wang, Z.},
	journal={Applied Mathematics and Computation},
	volume={428},
	pages={127061},
	year={2022},
	publisher={Elsevier}
}

@article{rosenzweig1963graphical,
	title={Graphical representation and stability conditions of predator-prey interactions},
	author={Rosenzweig, M.~L. and MacArthur, R.~H.},
	journal={The American Naturalist},
	volume={97},
	number={895},
	pages={209--223},
	year={1963},
	publisher={Science Press}
}

@article{sun2022dynamic,
	title={Dynamic analysis of a plant-water model with spatial diffusion},
	author={Sun, G.-Q. and Zhang, H.-T. and Song, Y.-L. and Li, L. and Jin, Z.},
	journal={Journal of Differential Equations},
	volume={329},
	pages={395--430},
	year={2022},
	publisher={Elsevier}
}

@article{sun2022impacts,
	title={Impacts of climate change on vegetation pattern: Mathematical modeling and data analysis},
	author={Sun, G.-Q. and Li, L. and Li, J. and Liu, C. and Wu, Y.-P. and Gao, S. and Wang, Z. and Feng, G.-L.},
	journal={Physics of Life Reviews},
	volume={43},
	pages={239--270},
	year={2022},
	publisher={Elsevier}
}

@article{oliveira2010modelling,
	title={Modelling disease introduction as biological control of invasive predators to preserve endangered prey},
	author={Oliveira, N.~M. and Hilker, F.~M.},
	journal={Bulletin of mathematical biology},
	volume={72},
	pages={444--468},
	year={2010},
	publisher={Springer}
}

@article{venturino2016ecoepidemiology,
	title={Ecoepidemiology: a more comprehensive view of population interactions},
	author={Venturino, E.},
	journal={Mathematical Modelling of Natural Phenomena},
	volume={11},
	number={1},
	pages={49--90},
	year={2016},
	publisher={EDP Sciences}
}

@article{bhattacharyya2006pest,
	title={Pest control through viral disease: mathematical modeling and analysis},
	author={Bhattacharyya, S. and Bhattacharya, D.~K.},
	journal={Journal of Theoretical Biology},
	volume={238},
	number={1},
	pages={177--197},
	year={2006},
	publisher={Elsevier}
}

@article{jana2013mathematical,
	title={A mathematical study of a prey--predator model in relevance to pest control},
	author={Jana, S. and Kar, T.~K.},
	journal={Nonlinear Dynamics},
	volume={74},
	pages={667--683},
	year={2013},
	publisher={Springer}
}

@article{chattopadhayay2002toxin,
	title={Toxin-producing plankton may act as a biological control for planktonic blooms—field study and mathematical modelling},
	author={Chattopadhayay, J. and Sarkar, R.~R. and Mandal, S.},
	journal={Journal of Theoretical Biology},
	volume={215},
	number={3},
	pages={333--344},
	year={2002},
	publisher={Elsevier}
}

@article{austin2002spatial,
	title={Spatial prediction of species distribution: an interface between ecological theory and statistical modelling},
	author={Austin, M.~P.},
	journal={Ecological modelling},
	volume={157},
	number={2-3},
	pages={101--118},
	year={2002},
	publisher={Elsevier}
}

@article{chen2011spatio,
	title={Spatio-temporal ecological models},
	author={Chen, Q. and Han, R. and Ye, F. and Li, W.},
	journal={Ecological Informatics},
	volume={6},
	number={1},
	pages={37--43},
	year={2011},
	publisher={Elsevier}
}

@article{hastings1990spatial,
	title={Spatial heterogeneity and ecological models},
	author={Hastings, A.},
	journal={Ecology},
	volume={71},
	number={2},
	pages={426--428},
	year={1990},
	publisher={JSTOR}
}

@article{sudakow2022knowledge,
	title={Knowledge gaps and missing links in understanding mass extinctions: Can mathematical modeling help?},
	author={Sudakow, I. and Myers, C. and Petrovskii, S. and Sumrall, C.~D and Witts, J.},
	journal={Physics of Life Reviews},
	volume={41},
	pages={22--57},
	year={2022},
	publisher={Elsevier}
}

@article{sekerci2020climate,
	title={Climate change forces plankton species to move to get rid of extinction: mathematical modeling approach},
	author={Sekerci, Y.},
	journal={The European Physical Journal Plus},
	volume={135},
	number={10},
	pages={794},
	year={2020},
	publisher={Springer Berlin Heidelberg}
}

\end{document}